\documentclass[twocolumn,showpacs,aps,prl]{revtex4}

\usepackage{graphicx}%
\usepackage{dcolumn}
\usepackage{amsmath}
\usepackage{latexsym}

\begin {document}

\title
{
Weighted Trade Network in a Model of Preferential Bipartite Transactions
}
\author
{
Abhijit Chakraborty$^{1}$ and S. S. Manna$^{1,2}$
}
\address
{
\begin {tabular}{c}
$^1$Satyendra Nath Bose National Centre for Basic Sciences,
Block-JD, Sector-III, Salt Lake, Kolkata-700098, India \\
$^2$Max-Planck-Institute f\"ur Physik Komplexer Systeme,
    N\"othnitzer Str. 38, D-01187 Dresden, Germany \\
\end{tabular}
}
\begin{abstract}

      Using a model of wealth distribution where traders are characterized by quenched random 
   saving propensities and trade among themselves by bipartite transactions, we mimic the enhanced 
   rates of trading of the rich by introducing the preferential selection rule using a pair of 
   continuously tunable parameters. The bipartite trading defines a growing trade network of 
   traders linked by their mutual trade relationships. With the preferential selection rule this 
   network appears to be highly heterogeneous characterized by the scale-free nodal degree and the 
   link weight distributions and presents signatures of non-trivial strength-degree correlations. 
   With detailed numerical simulations and using finite-size scaling analysis we present evidence 
   that the associated critical exponents are continuous functions of the tuning parameters. However 
   the wealth distribution has been observed to follow the well-known Pareto law robustly 
   for all positive values of the tuning parameters.
\end{abstract}
\pacs {
       89.65.Gh, 
       89.75.Hc, 
       89.75.Fb, 
       05.70.Jk  
}
\maketitle

\section {I Introduction}

      In a trading society different traders trade among themselves. Thus the wealth distribution
   of the society dynamically evolves through this trading process. In the simplest possible
   situation pairs of traders make economic transactions. Such a mutual interaction
   can be looked upon establishing a connection between them. Consequently one can define a trade 
   network where each trader is a node and a link is established between a pair of nodes when the 
   corresponding traders make a mutual trading. In this paper we study the growth and the structural 
   properties of a trade network in the framework of a well known model of wealth 
   distribution, namely the Kinetic Exchange Model (KEM) with quenched random saving propensities 
   \cite {Yakovenkoreview, ACandBKC}.

      Over the last decade tremendous amount of research efforts have been devoted to study the 
   structures, properties and functions of different real-world as well as theoretically defined 
   model networks \cite {Vesp}. The key characteristic features of these networks include their 
   small-world properties, which simply implies the existence of a very short global connectivity 
   even when the sizes of the networks are extremely large \cite {Watts}. Secondly, there are a 
   large class of networks that are extremely heterogeneous. Their heterogeneity are quantified 
   by their degree (number of links meeting at a node) distributions. Usually such networks are
   observed to have power law decay of their degree distributions and are called 
   Scale-free Networks \cite {Bara}. It has also been apparent that the links of many of these
   networks appear with a wide variation of strengths. In graph theoretic language the link
   strengths are called the `weights' in general. Such weighted networks have also been studied
   in the context of the passenger traffics of airport networks \cite {Guimera,Barrat}, 
   international trade networks (ITN) etc. \cite {Serrano,ITN}.
   
      More than a century ago Pareto proposed that the distribution 
   of wealth $x$ in a society to be ${\rm P}(x) \sim x^{-(\nu+1)}$ \cite {Pareto}. This form of distribution 
   is generally known as Pareto distribution for a value of $\nu \sim 1$ \cite {Wikipedia}. Pareto 
   suggested that $\nu = 1$ for the wealth distribution and it is known as the Pareto law.

      Application of the concepts of Statistical Physics to the wealth/income distribution in a society 
   goes back to 1931 when Saha and Srivastava had suggested that the form of the wealth distribution may 
   be similar to the Maxwell-Boltzmann distribution of molecular speeds in an ideal gas \cite {Saha, Sinha}.
   Over the last few years renewed attempts have been made using Statistical
   Physics methods. The main objective is to reproduce the recently collected individual income 
   tax data in different countries reflecting the wealth distributions. It has been observed that
   these data fit well to exponentially decaying functions for small wealths which however ends with
   power law tails in the large wealth regime. 

      Dr\u{a}gulescu and Yakovenko (DY) \cite {DY} modeled a bipartite trading between 
   two traders using the analogy of a pair of gas molecules interacting through an energy conserved elastic 
   collision in an ideal gas. Starting from an arbitrary distribution of individual wealths the system 
   evolves through a series of bipartite trades to arrive at a stationary state where the wealth distribution 
   assumes its time independent form. While the DY model produces only an exponentially decaying wealth 
   distribution, later modifications were suggested for its improvement. This class of models are now 
   referred as the KEMs \cite {Yakovenkoreview, ACandBKC}.

      In a KEM the society is considered as a collection of $N$ traders, each having a certain 
   amount of money equivalent to his wealth $x_i, \{i=1,N\}$ which he uses for mutual trades with other 
   traders. Generally all traders are initially given an equal amount of money ${\rm P}(x_i,t=0) = 
   \delta(x_i-a)$. The sequential time $t$ is the number of bipartite trades. A trade consists 
   of two parts, (i) a rule for the selection of a distinct pair of traders $i$ and $j$, $(i \ne j)$ and 
   (ii) a distribution rule for the random shuffling of their total money $x_i+x_j$ between 
   them. Different KEMs differ from one another either in the selection rule or in the distribution 
   rule or in both. In the DY model a pair of traders is selected with uniform probability. Their total 
   money is then randomly reshuffled between them. In the stationary state the wealth distribution is
   ${\rm P}(x) = \exp (-x/\langle x \rangle) / \langle x \rangle$ where the mean wealth $\langle x 
   \rangle = a$ is usually set at unity \cite {DY}. Quite naturally a trader invests a only a part of 
   his money in a trade and not his entire wealth. To incorporate this fact Chakraborti and Chakrabarti (CC) introduced a 
   saving propensity $\lambda$ \cite {CC} same for all traders. As a result the stationary state wealth 
   distribution gets modified to a distribution with a single maximum which approximately fits to a Gamma 
   distribution \cite {Patriarca}.

      In a second modification of the DY model Chatterjee, Chakrabarti and Manna (CCM) assigned 
   a quenched distribution of saving propensities $\{\lambda_i\}$ so that each trader is
   characterized by his own $\lambda_i$ value \cite {CCM}. Using the same selection 
   rule as in DY model, the total money invested by a pair of traders after saving has been 
   randomly shared between them. The system reaches a stationary state here as well but it sensitively 
   depends on the precise values of $\lambda_i$s. The wealth distribution in the stationary 
   state after averaging over different realizations of the quenched disorder $\{\lambda_i\}$ 
   yields a power law decay with a value of the Pareto exponent $\nu \approx 1$ \cite {CCM}. 
   However, subsequent detailed analyses have revealed that the CCM model has many interesting 
   features \cite {Kunal}. For example, in contrast to the DY and CC models, CCM model is not     
   ergodic. Therefore the wealth distribution is not self-averaging and the single trader
   wealth distribution is totally different from the over all wealth distribution of the whole 
   society. Consequently the individual saving propensity factor $\lambda_i$ plays the role 
   of an identification label that determines the economic strength of a member in the society \cite {Kunal}.
   In fact, the wealth of a trader fluctuates around a mean value which depends 
   very sensitively on the precise value of $\lambda_i$. Larger the value of $\lambda$ higher 
   is the mean wealth. Truly the wealth distribution averaged over many uncorrelated quenched 
   $\{\lambda_i\}$ sets is the convolution of the individual members' wealth distributions \cite {Kunal}.
   This overall distribution for the whole system exhibits Pareto law but not the individual 
   member distributions. The exponent $\nu=1$ has been found to be exactly equal to unity in
   \cite {Mohanty,BasuMohanty}.

      In section II we describe our modification of the CCM model using the preferential selection 
   rule. The wealth distribution of the modified CCM model has been described in section III. In 
   section IV we define the associated trade network in terms of its nodal degrees and link weights. 
   Section V presents the degree distribution and the weight distribution is discussed in section VI. 
   We summarize in section VII.

\section {II The Model}

      It is a general observation that in a society the rich traders invest much more in trade and 
   therefore take part in the trading process more frequently than the poor ones. To incorporate this 
   fact in the CCM model that rich traders are preferentially selected more frequently with higher 
   probabilities we introduce two parameters $\alpha$ and $\beta$ in general, both $\ge 0$, to tune 
   the preference different traders receive for their selections. We assume that the probability of 
   selection of a trader is directly proportional to the $\alpha$ and $\beta$-th power of its wealth. 
   Therefore a pair of traders $i$ and $j$ $(i \ne j)$ with money $x_i$ and $x_j$ are selected with 
   probabilities
\begin {equation}
\pi_i(t) \propto x_i(t)^{\alpha} {\hspace*{1.0 cm}} {\rm and} 
{\hspace*{1.0 cm}} \pi_j(t) \propto x_j(t)^{\beta}.
\end {equation}
   When $\alpha = \beta = 0$ we get the ordinary CCM. When they are non-zero the
   rich traders are selected with larger probabilities. More rich a trader, higher is the probability
   that it will be selected for trading.
   Once a pair of traders $i$ and $j$ is selected, they save $\lambda_i$ and $\lambda_j$ 
   fractions of their money and invest the rest amounts to the mutual trade. The total invested amount by
   both the traders is therefore
\begin {equation}
\delta_{ij}(t) = \bar {\lambda}_ix_i(t)+\bar {\lambda}_jx_j(t),
\end {equation}
   where $\bar {\lambda} = 1 - \lambda$. 
   This amount is then randomly divided into two parts and received by them randomly:
\begin {eqnarray}
& x_i(t+1)=\lambda_ix_i(t) +   \epsilon(t)\delta_{ij}(t) \nonumber \\
& x_j(t+1)=\lambda_jx_j(t) +\bar {\epsilon}(t)\delta_{ij}(t).
\end {eqnarray}
   where $\epsilon(t)$ is a freshly generated random fraction and $\bar {\epsilon} = 1 - \epsilon$. 

      It is essential that all measurements are done once the system attains the stationary state. 
   For that it is necessary that a number of transactions take place between every pair of traders, 
   only then the mean wealths of traders attain their stationary values and fluctuate around them. 
   Eqn. (1) states that for any $(\alpha,\beta) > 0$ the richest and the next rich are the most 
   probable pair and the poorest and the next poor are the least probable pair for selection. 
   Assuming the maximum wealth $x_{max} \sim N$ (with $\langle x \rangle$=1) and the minimal wealth 
   $x_{min} \sim 1/N$ the relaxation time can be estimated which is the typical time required for 
   the poorest pair to make a trade. The poorest is selected with a probability 
   $x^{\alpha}_{min}/\Sigma_i x^{\alpha}_i$. Approximating the denominator by its maximum value we 
   get $(x_{min}/x_{max})^{\alpha} \sim N^{-2\alpha}$. Similarly the probability for the next poorest 
   is $N^{-2\beta}$ and for the poorest pair as $N^{-2(\alpha+\beta)}$. Therefore the time required 
   for a trade between the poorest pair $ T_2 \sim N^{2(\alpha+\beta)}$ (see below) and the relaxation time is 
   several multiples of $T_2$. Thus for any $(\alpha,\beta) > 0$ the relaxation time grows very 
   rapidly with $N$.

\begin{figure}[top]
\begin{center}
\includegraphics[width=6.5cm]{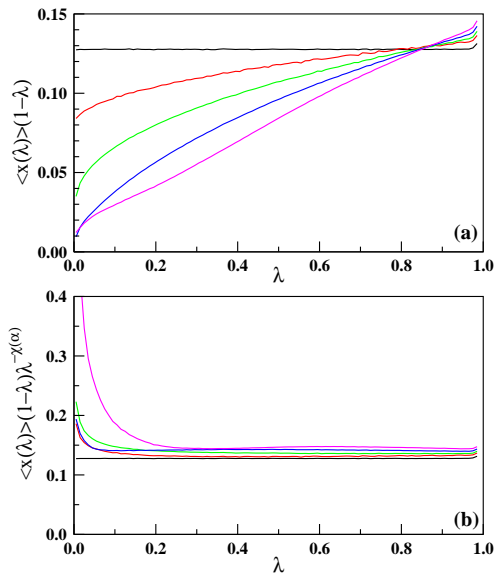}
\end{center}
\caption{
(Color online)
(a) Plots of $\langle x(\lambda) \rangle (1-\lambda)$ vs. $\lambda$ for $\alpha= \beta=0$ (black), 1/2 
(red), 1 (green), 3/2 (blue) and 2 (magenta) for $N$ = 1204 ($\alpha$ increases from top to bottom). 
(b) The product function $\langle x(\lambda) \rangle (1-\lambda) \lambda^{-\chi(\alpha)}$ is plotted
with $\lambda$ using $\chi(\alpha)$ = 0.15, 0.35, 0.57 and 0.80 for $\alpha$ = 1/2, 1, 3/2 and 2 respectively
using the same colors as in (a).
}
\end{figure}

      At the early stage rich traders at the top level quickly take part in the trading but gradually the 
   inclusion of relatively poor traders becomes increasingly slower. As a result the number of distinct 
   traders taking part in the trading process grows very slowly. Effectively this implies that the system
   passes through a very slow transient phase which is practically time independent. We call this state as 
   the ``quasi-stationary state (QSS)''. It is to be noted that in the following sections we present our 
   numerical results for large system sizes in the QSS only. To ensure that the system has indeed reached 
   the QSS in our simulations we keep track of the quantity $\Sigma_ix^2_i$ and collect the data only 
   after no appreciable change of its mean value is noticed. We mostly analyse the symmetric $\alpha=\beta$ 
   cases except for a few asymmetric cases.

\section {III Wealth Distribution}

      For CCM it was observed that the average money of a trader $\langle x(\lambda) \rangle$ with saving 
   propensity $\lambda$ diverges as $\lambda \to 1$ \cite {Kunal}. Later it was shown that the divergence 
   is like $\langle x_i \rangle (1-\lambda_i) = $ constant  \cite {Patriarca1,Mohanty}. This is simply 
   because had there been a trader with $\lambda=1$, he would not invest any money at all but always 
   receives a share of the investments of the other traders! As a result this trader will eventually grab 
   all the money of the society and, and this situation is similar to the phenomenon of condensation. 

\begin{figure}[top]
\begin{center}
\includegraphics[width=6.5cm]{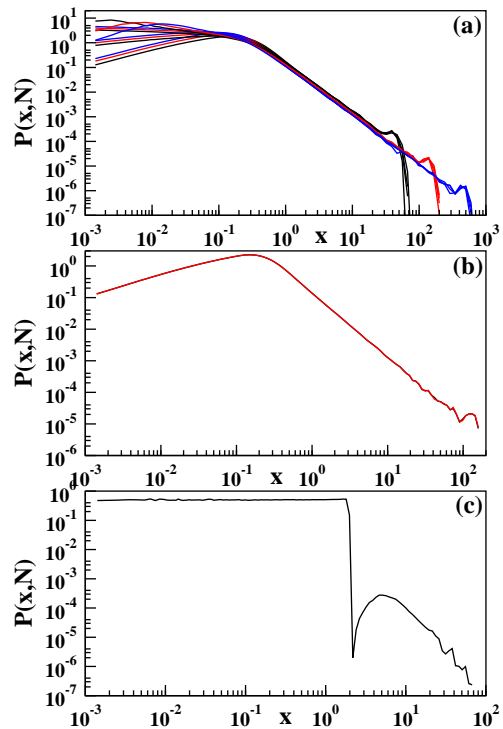}
\end{center}
\caption{(Color online)
(a) Wealth distribution $P(x,N)$ vs. $x$ for $\alpha=\beta$ = 0, 1/2, 1 and 2 and for $N$ = 256 (black), 1024 
(red) and 4096 (blue) ($N$ increases from left to right). The slopes of these curves yield $\nu = 1.00(3)$ 
consistent with the Pareto law.
(b) $P(x,N)$ for $(\alpha,\beta) = (\infty,0)$ (black) and $(0,\infty)$ (red) for $N$= 1024 which almost overlapped.
(c) $P(x,N)$ for $(\alpha,\beta) = (\infty,\infty)$, the distribution is uniform followed by a hump due to 
transactions between the richest and the next richest traders only.
}
\end{figure}

      In Fig. 1(a) we plot the quantity $\langle x(\lambda) \rangle (1-\lambda)$ vs. $\lambda$ for five 
   different values of $\alpha=\beta$ = 0, 1/2, 1, 3/2 and 2. For $\alpha=\beta=0$ we see the horizontal line 
   as observed in \cite {Patriarca1}. However for other $\alpha,\beta$ values the variations of the same 
   quantity are far from being uniform and are monotonically increasing with $\lambda$, their growth becoming
   increasingly faster with $\alpha$. Therefore we try multiplying this function by $\lambda^{-\chi(\alpha)}$
   where $\chi(\alpha)$ is a function of the parameter $\alpha$. In Fig. 1(b) we plot
   $\langle x(\lambda) \rangle (1-\lambda) \lambda^{-\chi(\alpha)}$ vs. $\lambda$ using the same data
   of Fig. 1(a) using $\chi(\alpha) =$ 0.15, 0.35, 0.57 and 0.80 for $\alpha=\beta=$ 1/2, 1, 3/2 and 2 respectively.
   We get nearly uniform variations between $\lambda=0.3$ and 1. We assume that
\begin {equation}
\langle x(\lambda) \rangle (1-\lambda) \lambda^{-\chi(\alpha)} = constant.
\end {equation}

\begin{figure}[top]
\begin{center}
\includegraphics[width=5.5cm]{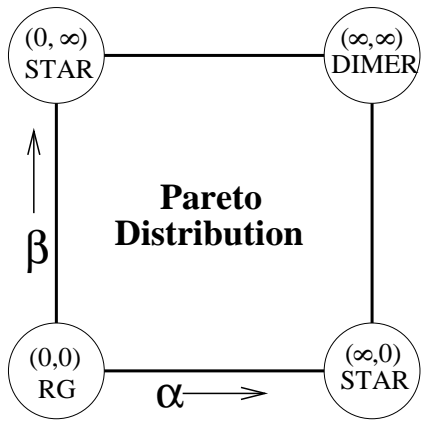}
\end{center}
\caption{
The phase diagram in the positive quadrant of the $(\alpha,\beta)$ plane. The Pareto law is valid in the 
entire region. The origin corresponds to the CCM model where the trade network is a random graph. At the
corners $(\infty,0)$ and $(0,\infty)$ the richest trader trade in every transaction, so that the network
is a star-like graph but the wealth distribution still follows Pareto law as shown in Fig. 2(b). However 
at the corner $(\infty,\infty)$ the trade takes place between the top two richest traders and the network
shrinks to a single dimer. The wealth distribution here is uniform followed by a hump as shown in Fig. 2(c). 
}
\end{figure}

      If the distribution of $\lambda$ values is denoted by $g(\lambda) = constant$, and since the wealth 
   $x$ and the saving propensity $\lambda$ are the two mutually dependent variables associated with the 
   same trader, their probability distributions follow the relation \cite {Mohanty}
\begin {equation}
P(x)dx=g(\lambda)d\lambda.
\end {equation}
   Differentiating Eqn. (4) with $\lambda$ one can find out the derivative $d\lambda/dx$ and substituting in Eqn. (5) 
   one gets
\begin {equation}
P(x) = \frac{C}{x^2}[\lambda^{-\chi}+(1-\lambda)\chi\lambda^{-\chi-1}]^{-1}.
\end {equation}
   For this equation we see that for large $\lambda$ the term within $[..]$ is of the order of unity.
   Therefore in this range $P(x) \sim x^{-2}$ as in the Pareto law. This is an indication that 
   even for $(\alpha,\beta) > 0$, Pareto law holds good and in the following we present numerical evidence
   in support of that.

      The system is prepared by assigning uniformly distributed random fractions for the saving
   propensities $\lambda_i$ to all $N$ traders. Here $\lambda_i$s are quenched variables and
   therefore they remain fixed during the subsequent time evolution of the trading system.
   Consequently all observable that we measured are averaged over different uncorrelated
   sets of the $\{\lambda_i\}$ values. While assigning
   the $\lambda$ values we first draw $N$ uniformly distributed random fractions, but then to avoid the situation
   when $\lambda_{max}$ is very close to unity by chance we scale them proportionately so that
   $\lambda_{max} = 1-1/N$ in every $\{\lambda_i\}$ set. First a pair of values for ($\alpha,\beta$)
   is selected. Two types of initial wealth distributions are used: (i) $x_i=1$ for all $i$ and (ii) $x_i$s are uniformly 
   distributed random numbers with $\langle x \rangle$ = constant. The sequence of bipartite trading begins by 
   randomly selecting pairs of traders
   using Eqn. 1. Once a pair is selected, their total individual invested amount $\delta_{ij}$ is
   calculated using Eqn. 2 and this amount is shared again between them using Eqn. 3. This constitute
   a single bipartite trading and the dynamics is followed over a large number of such trading events.

      The wealth distribution changes with time from the initial distribution to more and more flat distribution. 
   After a certain time the system passes through the quasi stationary state when no appreciable change in the 
   wealth distribution is observed. It is also observed that the distribution is robust with respect to the precise values of 
   the parameters $\alpha$ and $\beta$ used. In Fig. 2(a) the wealth distribution $P(x,N)$ has been plotted with 
   $x$ for four sets of parameter values namely, $\alpha = \beta$ = 0, 1/2, 1 and 3/2 and for three system sizes 
   $N$ = 256, 1024 and 4096. Apart from slight fluctuations the four curves for a given system size nearly overlap 
   on one another. On a double logarithmic scale the slopes of the curves give an average estimate for $\nu = 1.00(3)$ 
   consistent with the Pareto law as observed in the CCM \cite {CCM}. This indicates that the wealth distribution 
   is robust with respect to the parameter values 
   in this region. The non-zero values of $\alpha$ and $\beta$ only controls the frequencies with which different 
   traders are called for trading.
\begin{figure}[top]
\begin{center}
\includegraphics[width=6.5cm]{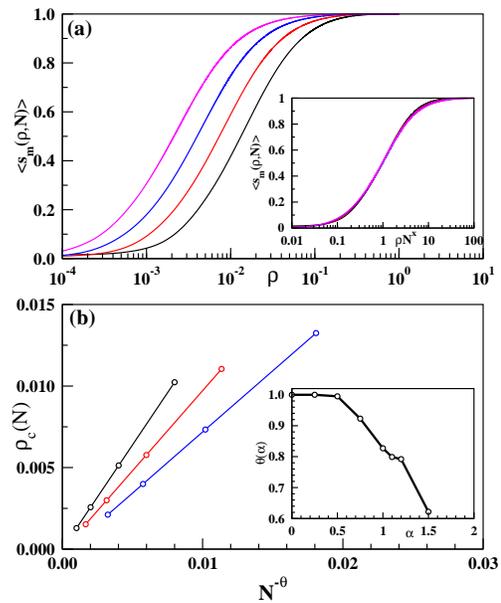}
\end{center}
\caption{(Color online)
Growth of the trade network.
(a) Plot of the average size of the giant component $\langle s_m(\rho,N) \rangle$ with the link 
density $\rho$, for $\alpha=\beta=1$ and for $N$ = 128 (black), 256 (red), 512 (blue) and 1024 
(pink), ($N$ increases from right to left). The inset shows a data collapse of the same plots with
$\rho N^{\theta}$, and ${\theta}=0.88$.
(b) The percolation link density $\rho_c(N)$ is plotted with $N^{-\theta}$ where $\theta$ = 0.88, 0.92 and 1 for 
$\alpha=\beta=1/2, 3/4$ and 1 respectively. The inset plots $\theta(\alpha)$ with $\alpha$. 
}
\end{figure}
\begin{figure}[top]
\begin{center}
\includegraphics[width=6.5cm]{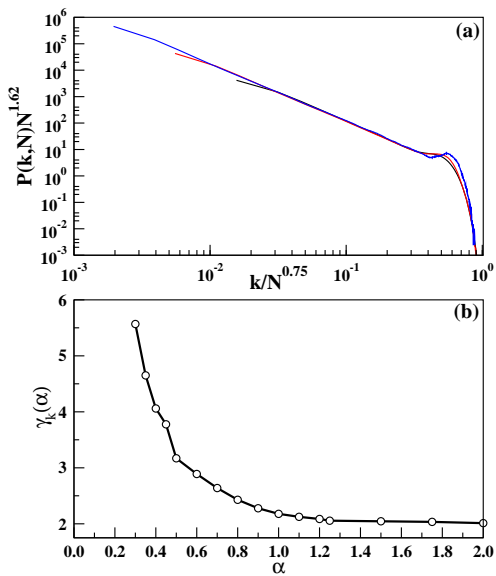}
\end{center}
\caption{(Color online)
(a) The finite-size scaling of the degree distributions $P(k,N)$ for $\alpha=\beta=1$, for $N$ = 256 
(black), 1024 (red) and 4096 (blue), and for $\langle k \rangle$ = 1. Direct measurement of slopes give
$\gamma_k(1) = 2.18(3)$. The best data collapse corresponds to $\eta_k(1) = 1.62$ and $\zeta_k(1) = 0.75$ 
giving $\gamma_k(1) = 2.16(3)$. 
(b) The plot of $\gamma_k(\alpha)$ vs. $\alpha$.
}
\end{figure}

      Next we consider the case when one of the two parameters $(\alpha,\beta)$ is infinity and the other 
   one is zero. If $\alpha=\infty$ the richest trader is always selected as the first trader. The other 
   trader is selected among the other $N-1$ traders with uniform probability. As shown in Fig. 2(b) here 
   also we see that the Pareto law holds good. For $\alpha=\infty$ and for finite $\beta$
   first the richest trader is selected and then the second trader is selected with probability 
   $\propto x_j^{\beta}$. We observe numerically that here also Pareto law works very well.

      However the situation is very different when both $(\alpha,\beta)$ take very large values. In this 
   situation almost always only the rich traders are called for transactions. The system passes through an 
   extremely long QSS and the number of traders taking part in trade does not increase at all. For example 
   in the limiting case of $(\alpha,\beta)$ = $(\infty,\infty)$ it implies that always only the richest and 
   the next richest traders are selected for transactions with probability one but not any other trader. If 
   their wealths are very high then the trading will be limited only between them. Therefore the wealth 
   distribution for the single ${\lambda_i}$ set has two very high peaks and wealths of all other traders 
   are small and uniformly distributed. Consequently the quench averaged wealth distribution is uniform 
   throughout followed by a hump at the highest value of wealth (Fig. 2(c)). A systematic analysis with 
   many different $(\alpha,\beta)$ pairs leads us to conclude that Pareto law holds good in the positive 
   quadrant of the entire $(\alpha,\beta)$ plane. 

      In Fig. 3 we exhibit this behavior in the positive quadrant of the $(\alpha,\beta)$ plane where 
   Pareto law is valid and the limiting points are marked by circles with their characteristics. The 
   origin at $(\alpha=0,\beta=0)$ represents the CCM model where traders are selected randomly with uniform 
   probabilities. As explained below, the trade network corresponding to this point is a random graph (RG).
   As explained before that at the two corners $(\infty,0)$ and $(0,\infty)$ the richest trader always 
   participates in every transaction. Therefore the corresponding trade networks have star-like structures.
   In the last corner of $(\infty,\infty)$ the trade takes place only between the richest and the next rich
   traders and therefore the graph reduced to a single dimer only.

\begin{figure}[top]
\begin{center}
\includegraphics[width=6.5cm]{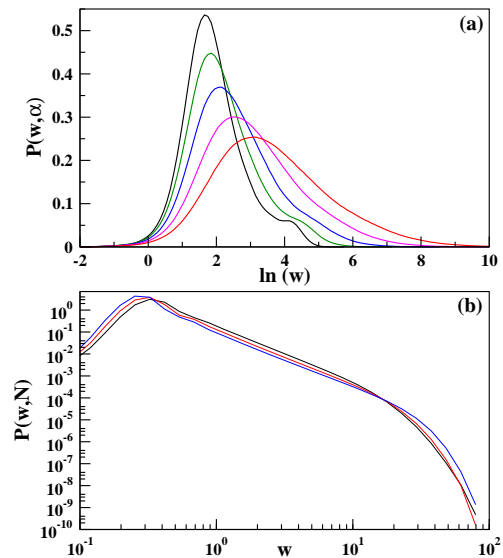}
\end{center}
\caption{
(Color online)
Probability distribution $P(w,\alpha)$ of the link weights. (a) Plots for the $N$-clique graphs with $N$ =64 
and for $\alpha$= 0 (black), 1/4 (green), 1/2 (blue), 3/4 (magenta) and 1 (red) and $\beta$ =1 always. (b) 
Plots for $\langle k \rangle$ = 5 with $\alpha = \beta = 1$ and for the system sizes 128 (black), 256 (red) 
and 512 (blue). Direct measurement of slopes gives 2.52, 2.53 and 2.51 respectively.
}
\end{figure}

\section {IV The Trade Network}

      One can associate a network with this trading system. Each trader is a node of the network. Initially 
   the network has only $N$ nodes but no links. First the system is allowed to reach the QSS and then the 
   network starts growing. Every time a pair of traders makes a trade for the first time, a link is 
   introduced between their nodes. There after no further link is added between them irrespective of their 
   subsequent trades and they remain connected with a single link. As the system evolves more and more 
   new traders take part in the trading dynamics and consequently the number of links grow in the network.
   For $\alpha=\beta=0$ the growth of the network is exactly the same as that of the random graph, however 
   it is much different when $(\alpha,\beta) > 0$. Since the rich nodes are preferentially selected they 
   acquire links at a faster rate than the poor nodes. The degree $k_i$ of the node $i$ is the number of 
   distinct traders with whom the $i$-th trader has ever traded. The dynamics is continued for a certain
   time $T$ till the average degree $\langle k \rangle$ of a node reaches a specific pre-assigned value.

      In general there are two characteristic time scales involved. At the early stage the network grows with 
   multiple components with different sizes. At time $T_1$ the growing network becomes a single component 
   connected graph. A second time scale is $T_2$ when the whole network is a $N$-clique graph in which each 
   node is linked to all others, which means each trader has traded at least once with all others. Unlike 
   random graphs the growth of the network is highly heterogeneous and the rich traders have much larger degrees 
   than the poor traders. Since poor traders are selected with low probabilities they take much longer times 
   to be a part of the network. Consequently $T_1$ is found to be very large and of the same order as $T_2$. 
   Numerically it is easier to calculate $T_2$, one keeps track of the number of distinct links and stops only 
   when this number becomes just equal to $[N(N-1)]/2$. On the other hand to calculate $T_1$ one follows the 
   growth of the giant component and stops when the giant component covers all $N$ nodes. A Hoshen-Kopelman 
   cluster counting algorithm \cite {Hoshen} is used to estimate the size of the giant component. For the 
   ordinary CCM with $\alpha = \beta = 0$ since both traders are chosen with uniform probability, the generated 
   graph is a simple Erd\H os-R\'enyi random graph characterized by a Poissonian degree distribution \cite {Erdos}.

\begin{figure}[top]
\begin{center}
\includegraphics[width=6.5cm]{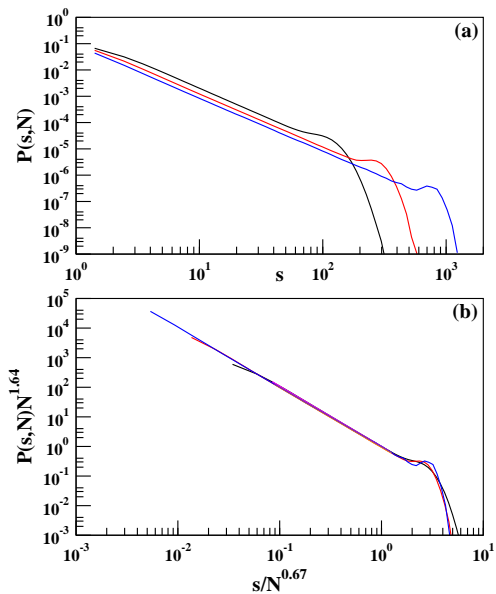}
\end{center}
\caption{
(Color online)
(a) Nodal strength distributions $P(s,N)$ with strength $s$ with $\alpha=\beta=1$, for $N$ = 256, 1024 and 4096 
and for $\langle k \rangle$ = 5. Direct measurement of the slopes of these curves gives $\gamma_s = 2.50(5)$.
(b) The Finite-size scaling analysis of the data in (a) gives $\eta_s = 1.64$ and $\zeta_s = 0.67$ estimating 
the value of $\gamma_s =\eta_s/\zeta_s= 2.45(5)$.
}
\end{figure}

      The growth of the giant component is studied with increasing number of links $n$ in the in the trade 
   network. The average fraction of nodes in the giant component is denoted by $\langle s_m(\rho,N) \rangle$ 
   which is the order parameter in this percolation problem. This has been plotted in Fig. 4(a) using a 
   semi-log scale with link density $\rho = n / [N(N-1)]/2$ in the network. Four curves shown in this figure 
   correspond to $N$ = 128, 256, 512 and 1024 for $\alpha=\beta=1$, the system size increasing from right to 
   left. The inset shows that a data collapse can be obtained by scaling the $\rho$ axis by a factor $N^{\theta}$ 
   with $\theta=0.88$. The critical density of percolation transition $\rho_c(N)$ is defined as that particular 
   value of $\rho$ for which the average size of the giant component $\langle s_m(\rho,N) \rangle = 1/2$. In 
   Fig. 4(b) we show that how the critical percolation threshold $\rho_c(N)$ depends on $N$ by plotting it with 
   $N^{-\theta}$ for $\alpha=\beta=$ 1/2, 3/4 and 1. It has been observed that the exponent $\theta(\alpha)$ 
   is dependent on $\alpha$ in general and in the inset of this figure we plot $\theta(\alpha)$ vs. $\alpha$. 
   We see that for $\alpha \le 1/2$, $x(\alpha)=1$ but for $\alpha > 1/2$, $x(\alpha)$ gradually decreases. 
   For Erd\H os-R\'enyi random graphs it is known that $x=1$ and therefore this result gives an indication that 
   the trade network seems to be different from random graphs for $\alpha = \beta > 1/2$.
   
\begin{figure}[top]
\begin{center}
\includegraphics[width=6.5cm]{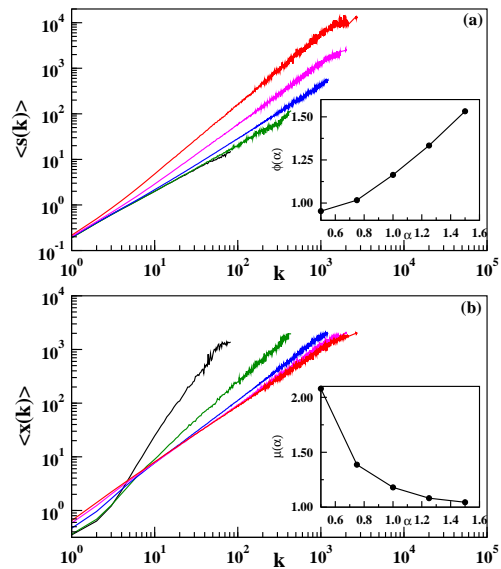}
\end{center}
\caption{
(Color online)
(a) Plot of nodal strength $\langle s(k) \rangle$ with degree $k$ for
$N=2^{14}$ and for $\alpha=\beta=1/2$ (black), 3/4 (green), 1 (blue), 5/4 (magenta) and 3/2 (red)
(from bottom to top). The slopes of these plots estimate $\phi(\alpha)$ shown in the inset.
(b) Plot of average wealth $\langle x(k) \rangle$ with degree $k$ for
$N=2^{14}$ and for $\alpha=\beta=1/2$ (black), 3/4 (green), 1 (blue), 5/4 (magenta) and 3/2 (red)
(from bottom to top). The slopes of these plots estimate $\mu(\alpha)$ shown in the inset.
}
\end{figure}

\section {V Degree Distribution}

      The degree distribution has been studied similar to random graphs. We keep track of 
   the average degree $\langle k \rangle$ of the network which is related to the number of links $n$ 
   of the network as $n = \langle k \rangle \frac{N}{2}$. First the degree distribution has been studied 
   for $\langle k \rangle$ = 1 and for different system sizes. For an assigned set of values of
   $(\alpha, \beta)$, for a given set of values for the saving propensities $\{\lambda_i\}$ and for a 
   specific value of $N$ once $\langle k \rangle$ = 1 is reached we calculate the degree distribution 
   considering all components of the network on the same footing. The network is then refreshed by 
   removing all links and a second network is 
   constructed and so on. The dynamics is continued for the same values of the parameters and the same set of 
   $\{\lambda_i\}$s till a large number of networks are generated and their mean degree distribution
   is calculated. The entire dynamical process is then repeated with another uncorrelated set of $\{\lambda_i\}$s 
   and the degree distribution has been averaged over many such sets.

      In Fig. 5(a) we show the finite-size scaling plot of the average degree distribution $P(k,N)$ vs. $k$ 
   for $\alpha = \beta = 1$ 
   and for $N$ = 256, 1024 and 4096. On a double logarithmic scale all three curves show quite long scaling 
   regions followed by humps before the cut-off sizes of the degree distributions. The cut-off sizes of the 
   distributions shifts to the larger values of $k$ approximately by equal amounts on the double-log scale 
   when the system size has been enhanced by the same factor. From the direct measurement of slope in the scaling
   region we estimate $\gamma(1) = 2.18(3)$. Almost the entire degree distribution 
   obeys nicely the usual finite-size scaling analysis and an excellent collapse of the data is observed
   confirming the validity of the following scaling form:
\begin {equation}
P(k,N) \propto N^{-\eta(\alpha)}{\cal G}[k/N^{\zeta(\alpha)}]
\end {equation}
   where the scaling function ${\cal G}(y)$ has its usual forms like, ${\cal G}(y) \sim y^{-\gamma(\alpha)}$ 
   as $y \to 0$ and ${\cal G}(y)$ approaches zero very fast for $y >> 1$. This is satisfied only when 
   $\gamma(\alpha) = \zeta(\alpha)/\eta(\alpha)$. The exponents $\eta(\alpha)$ and $\zeta(\alpha)$
   fully characterize the scaling of $P(k,N)$ in this case. To check the validity of the equation we attempted
   a data collapse by plotting $P(k,N) N^{\eta(1)}$ vs. $k/N^{\zeta(1)}$ by tuning the values
   of $\eta(1)$ and $\zeta(1)$. The values obtained for best data collapse are
   $\eta_k(1)=1.62$ and $\zeta_k(1)=0.75$ implying that in the infinite size limit $P(k,\infty) \sim k^{-\gamma(1)}$ 
   with $\gamma(1)$ = 2.16(3). 
   Tuning $\alpha$ and $\beta$ to other values it is observed that the degree distribution 
   exponent $\gamma$ does depend on these two parameters. In Fig. 5(b) we show a plot of $\gamma_k(\alpha)$
   with $\alpha$ which decreases to $\approx 2$ at $\alpha=2$.

\section {VI The Weighted Network}

      Within a certain time $T$ a large number of bipartite trades take place between any arbitrary pair of traders. 
   The total sum of the amounts $\delta_{ij}$ invested in all trades between the traders $i$ and $j$ in time $T$ is 
   defined as the total volume of trade $w_{ij} = \Sigma^T \delta_{ij}$. Therefore $w_{ij}$ is regarded as 
   the weight of the link $(ij)$. The magnitudes of weights associated with the links of the trade network are again 
   found to be highly heterogeneous. This is primarily because within a certain time $T$ a rich pair of traders trade 
   many more times than a rich-poor or a poor-poor pair. In addition the invested amounts depend on the mean wealths 
   $\langle x_i \rangle$ of the traders involved as well as their saving propensity factors $\lambda_i$. The 
   probability distribution $P(w,N)$ of the link weights are calculated when the average degree $\langle k \rangle$ 
   reaches a specific pre-assigned value. As before, this distribution has also been averaged over many weighted 
   networks for one $\{\lambda_i\}$ set and then further averaged over many uncorrelated $\{\lambda_i\}$ sets.
   
      First we studied 
   the case when the trade networks is a $N$-clique graph, i.e., when each trader has traded with all other traders 
   at least once. Here each node has same degree i.e., $P(k)=\delta(k-(N-1))$ and $\langle k \rangle = N-1$. The 
   required time $T_2$ increases rapidly with $N$ as described in section II and we could study small system size
   $N$ = 64 only. The distribution has a very long tail and therefore we used a lin-log scale for plotting. In Fig. 
   6(a) we show the plots of $P(w,\alpha)$ with $\ln (w)$ for different values of $\alpha$ = 0, 1/4, 1/2, 3/4 and 1
   and $\beta=1$. Each curve is asymmetric and has a single maximum. The position of the peak shifts towards larger 
   values of $\ln(w)$ as $\alpha$ increases. If Fig. 6(b) a similar plot has been shown for $\langle k \rangle = 5$
   for three network sizes $N$ = 128, 256 and 512 and for $\alpha=\beta=1$. On a double-logarithmic scale each curve 
   has a considerable 
   straight portion. This indicates a power law decay like $P(w,N) \propto w^{-\gamma_w}$. The 
   corresponding slopes give estimates for the exponent $\gamma_w$ as 2.52, 2.53 and 2.51 for the three system sizes
   respectively, so that on the average $\gamma_w = 2.52(3)$.

      The strength of a node $s_i = \Sigma_j w_{ij}$ where $j$ runs over all neighbors $k_i$ of $i$, is a measure 
   of the total volume of trade handled by the $i$-th node. Nodal strengths varies over different
   nodes over a wide range. We first study the probability distribution of nodal strengths. In Fig. 7(a) the
   strength distribution $P(s,N)$ has been plotted for the average degree $\langle k \rangle =5$, for $\alpha=\beta=1$
   and for the network sizes $256, 1024$ and 4096. Extended scaling regions at the intermediate regions of the curves 
   indicate that $P(s,N)$ also follows a power-law decay function $P(s,N) \sim s^{-\gamma_s}$ in the limit
   of $N \to \infty$. Direct measurements gives an estimate of $\gamma_s(1) \approx 2.5$. In Fig. 7(b) we try a
   similar finite size scaling of the same data giving $\eta_s(1)=1.64$ and $\zeta_s(1)=0.67$ giving $\gamma_s(1)=2.45(5)$.

      Quite often weighted networks have non-linear strength-degree relations reflecting the presence of non-trivial 
   correlations, example of such networks are the airport networks and the international trade network. For a network 
   where the link weights are randomly distributed, the $\langle s(k) \rangle$ grows linearly with $k$. However a 
   non-linear growth like $\langle s(k) \rangle \sim k^{\phi}$ with $\phi > 1$, exhibits the presence of 
   non-trivial correlations. In Fig. 8(a) we plot the variation of 
   $\langle s(k) \rangle$ vs. $k$ for a system size $N$ = 16384 and for different values of
   $\alpha=\beta=1/2$ (black), 3/4 (green), 1 (blue), 5/4 (magenta) and 3/2 (red)
   (from bottom to top). The slopes of these plots give estimates for the exponent $\phi(\alpha)$ which
   gradually increased with $\alpha$ and the variation has been plotted in the inset.
   In the same context we also studied how the mean wealth of a trader depends on its degree. The mean
   wealth of a trader $\langle x(k) \rangle$ has been plotted in Fig. 8(b) with its degree $k$ for
   the same system sizes as in Fig. 8(a) and for the same values of parameters. A power law growth has been
   observed for all values of $\alpha$: $\langle x(k) \rangle \sim k^{\mu(\alpha)}$. 
   The slopes of these plots give estimates for the exponent $\mu(\alpha)$ which
   has been plotted in the inset of Fig. 8(b).
   
\section {IV Summary}

      To summarize we have studied the different structural properties of a trade network associated
   with the dynamical evolution of a model of wealth distribution with quenched saving propensities.
   In this model distinguishable traders make preferential bipartite trades among themselves and in 
   this way create links. They are selected for trade preferentially using a pair of continuously
   tunable parameters, where the rich traders are picked up more frequently for trade than poor traders. 
   This creates huge heterogeneity in the system which has been reflected in the power-law distributions 
   of the nodal degree and the link weight distributions measuring the volumes of trade. We present
   numerical evidence that the associated individual wealth distribution follows the well known Pareto 
   law robustly for all positive values of the parameters.

      We thankfully acknowledge P. K. Mohanty, A. Chatterjee and B. K. Chakrabarti for discussion and critical reading of the manuscript.

\leftline {E-mail: manna@bose.res.in}

\end {document}